\documentclass[12pt]{amsart}
\usepackage{macrosavb}
\usepackage{geometry}                % See geometry.pdf to learn the layout options. There are lots.
\usepackage[parfill]{parskip}    % Activate to begin paragraphs with an empty line rather than an indent
\usepackage{graphicx}
\usepackage{amssymb}
\usepackage{epstopdf}
\usepackage[T1]{fontenc}

\title{Théorie de Galois effective : aide mémoire}
\author{Annick Valibouze}
\address{UPMC, 4, place Jussieu, 75252 Paris Cedex 05}
\email{annick.valibouze@upmc.fr \quad www-spiral.lip6.fr/\~{}avb/}

\date{\today}

\begin{document}
%\makeheadbox
\vspace{0.5in}
\maketitle
%${\boldsymbol a}$ 
%\Large
Cet article recense de nombreux résultats obtenus dans diverses publications relevant de la théorie de Galois effective ; en particulier, les formules et théorèmes sur les idéaux galoisiens peuvent s'exprimer de diverses fa{\c c}ons et sont parfois ``redécouverts'' par de triviales reformalisations. L'éparpillement complique également leur utilisation rapide.

Les idéaux galoisiens (dits alors de Galois) apparurent tout d'abord dans un 
support de cours et d'encadrement doctoral en théorie Galois effective (voir \cite{VALIBOUZE:1995:CEL-00403452:1}) ; ce support comportant un nombre important de résultats nouveaux servit également de document de travail au projet Galois du GDR (puis de l'UMS) MEDICIS du CNRS. 

Cet article a comme double objectif que 1) ne soient pas redécouverts des résultats déjà connus et 2) de les retrouver rapidement.

 \section{Données}
\begin{itemize}
\item $k$ un corps , $\ck$ une clôture algébrique de $k$
%\item $x_1,\ldots ,x_n, x$ variables indépendantes sur $k$
\item $\seq{x}{n},x$ variables indépendantes sur $k$
\item $f$ polynôme en $x$ de degré $n$ à coefficients dans $k$ %de racines distinctes
\item $\nracf$ les $n$ racines de $f$ dans $\ck$ :
$$ f = a_n\prod_{i=1}^n (x-\alpha_i) = a_nx^n+a_{n-1}x^{n-1}+\cdots + a_0$$
\end{itemize}

\section{Notations générales}
\begin{itemize}
\item $\SY_n$ : groupe symétrique de degré $n$.
\item $I_n$ : sous-groupe identité de $\SY_n$
\item $A_n$ : sous-groupe alterné de $\SY_n$
\item $\racf :=(\nracf)$ , $\varx=(x_1,\ldots x_n)$, $\coef=(a_1,\ldots ,a_n)$, ${\uplet{i}}=(i_1,\ldots ,i_n)$, etc $\ldots$
\item $\varx^{\uplet{i}} := x_1^{i_1}x_2^{i_2}\cdots x_n^{i_n}$,  etc $\ldots$
\item $$ pol(\uplet{y}):=\prod_{i=1}^n (x-y_i)$$
\end{itemize}

\section{Groupes}
{\bf 3.1 Généralités classiques} 

$G,H \in \SY_n$ 
\begin{itemize}
\item $G < H$ : $G$ sous-groupe de $H$
\item $G \subset H$ : $G$ sous-ensemble de $H$
\item $GH=\{gh \mid \; g \in G \; h\in H\}$
\item Hyp. $G < H$ ; $G_1, \ldots ,G_s$ sont {\it les classes à droite (resp. gauche) de $H$ modulo $G$} si pour $i=1,\ldots ,s$, $G_i=G\tau_i$ (resp. $\sigma_iG$) où $\tau_i \in H$ (resp. $\sigma_i \in H$), et $H$ est l'union disjointe des $G_i$ :
$$H = G_1 +\cdots + G_s$$
\item $G\backslash H := \{\tau_1,\ldots ,\tau_s\}$ est une {\it transversale à droite} de $H$ modulo $G$.
\item $H/G:= \{\sigma_1,\ldots ,\sigma_s\}$ est une {\it transversale à gauche} de $H$ modulo $G$.
\item $[H:G]:=s$,  l'{\it indice} de $G$ dans $H$
\item $G$ est un sous-groupe {\it distingué} ({\it normal})  de $H$ si $G < H$ et
$$(\forall \tau \in H) \; G=\tau G \tau^{-1}$$
(i.e. les classes à droite et à gauche sont identiques)
\item $G^\tau :=\tau G \tau^{-1}$
\item $G$ sous-groupe distingué de $H$ ssi la tranversale (droite et gauche) de $H$ modulo $G$ forme un sous-groupe de $H$  noté $H/G$ (=$G\backslash H$)
\end{itemize}

%{\bf Définition}
\begin{itemize}
\item {\it Action (à gauche)} de $G$, un groupe, sur un ensemble non vide $E$, toute opération notée $ \star$ :
    \begin{eqnarray*}
       \SY_n \times E \longrightarrow& E , \quad (\sigma,x)  \mapsto \sigma\star x
       \end{eqnarray*}       
vérifiant les axiomes suivants :
\begin{enumerate}
\item $(\forall x \in E)$, $e_G\star x=x$ , où $e_G$ est l'élément neutre de $G$
\item $(\forall x \in E) \, ,(\forall \sigma,\tau \in \SY)$, $\sigma\star (\tau\star x)=(\sigma\tau)\star x$
\end{enumerate}
\item $x \in E$, $\sigma \in G$ : 
\begin{eqnarray*}
\sigma\star E &:=&\{\sigma\star x \mid x \in E\} \\
G\star x &:=&\{\sigma \star x \mid \sigma \in G\} \quad \text{\it orbite de $x$ sous l'action de $G$}\\
G\star E & := & \{G\star x \mid x \in E\} = \{\sigma \star E \mid \sigma \in G\}\\
\Stab_{G}{x} &:= & \{\sigma \in G \mid \sigma\star x = x\} \quad \text{\it stabilisateur de $x$ dans $G$}
 \end{eqnarray*}       

\end{itemize}

{\bf 3.2 Actions particulières} 

Soit $\sigma \in \SY_n$.
\begin{itemize}
\item $\SY_n$ agit naturellement sur $E:=\{1,\ldots ,n\}$ comme groupe de permutations : 
 \begin{eqnarray*}
       \SY_n \times E \longrightarrow& E , \quad (\sigma,j)  \mapsto \sigma(j) = i_j \quad \text{si } \sigma =\left (\begin{array} {c} 1,\ldots ,n\\ i_1,\ldots ,i_n \end{array} \right ) 
       \end{eqnarray*}       
\item $\SY_n$ agit sur les $n$-uplets : 
$$\sigma *\uplet{y}:=(y_{\sigma(1)},\ldots ,x_{\sigma(n)}) $$
\item $\SY_n$ agit sur les monômes : %$\sigma\in \SY_n$, $\varx^\uplet{i} \in \avx$ 
$$\sigma.\varx ^{\uplet{i}}:= 
(\sigma *\varx) ^{\uplet{i}}=x_{\sigma(1)}^{i_1}x_{\sigma(2)}^{i_2}\cdots x_{\sigma(n)}^{i_n} $$
et par extension $\SY_n$ agit sur $\cvx$.
\item (plus tard) $\Gal_k(\racf)$ agit sur $k(\racf)$ : \\
$\Theta \in \avx$, $\theta=\Theta(\nracinesf) \in k(\racf)$, 
$\tau \in \Gal_k(\racf)$, 
$$\beta^\tau= \Theta(\alpha_{\tau(1)},\ldots ,\alpha_{\tau(n)})$$
Rq : $\Gal_k(\varx)=\SY_n$ agit sur $\cvx$; on pourrait noter $p^\sigma:=\sigma.p$, $p\in \cvx$
\item (plus tard) $\Gal_k(f):=\Aut_k(k(\racf))$  agit sur $k(\racf)$ : 
$$\Gal_k(f) \times k(\alpha) \longrightarrow k(\alpha), \quad  (\phi ,\beta) \mapsto \phi(\beta)$$
%\end{itemize}
%{\bf Notations actions} 
%\begin{itemize}
\item Notation : $r \in \avx$, $$\sigma.r (\racf):=(\sigma.r)(\racf)=r(\sigma *\racf)$$
\item $(\forall \sigma,\tau \in \SY_n)$, $(\forall r \in \avx)$,
$$\tau\sigma.r(\racf)=r(\tau\sigma * \racf)=\sigma.r(\tau*\racf)$$
(i.e. permuter d'abord $r$ avec $\sigma$ puis évaluer en $\tau*\racf$)
\end{itemize}

{\bf 3.3 Matrices des groupes et de partitions}\\

Soient $G,H < L< \SY_n$. Soit ${\mathcal O}$ l'ensemble des $G$-orbites de $L$ modulo $H$. Pour tout $O \in {\mathcal O}$ de cardinal $s$, on note $Gr(G,O)$, la représentation symétrique dans $\SY_s$ de l'action à gauche de $G$ sur $O$ ;
soit le vecteur de groupes
$$Gr_L(G,H):=[Gr(G,O) \mid O \in {\mathcal O}]$$
\begin{itemize}
\item $Gr_L(G,H)$ ne dépend pas de la classe de conjugaison de $G$ et $H$ dans $\SY_n$
\item {\it ${\mathcal G}(L)$ matrice des groupes de $L$}\\
elle est indicée en colonne et en ligne par les classes de conjugaison des sous-groupes de $S_n$
- Soient $G,H$ deux sous-groupes de $S_n$ ; à l'intersection de la ligne de la classe de $H$ et de la colonne de celle de $G$ se trouve le vecteur de groupes $Gr_L(G,H)$.

\item {\it ${\mathcal P}(L)$ matrice des partitions de $L$}\\
c'est la matrice déduite de ${\mathcal G}(L)$ en remplaçant les groupes par leur degré respectif.
\item Les lignes de la matrice de partitions (et donc de groupes) sont toutes distinctes.
\end{itemize}

\section{Idéaux galoisiens}

{\bf 4.1 Généralités classiques} 

$I$ idéal de $\avx$, $V \subset \ck^n$
\begin{itemize}
\item idéal définit par $V$ dans $\avx$ : 
$$\Id_{\avx}(V) :=\{r \in \avx \mid (\forall \uplet{b} \in V)\; r(\uplet{b}) = 0 \}$$
(par défaut $\Id(V):=\Id_{\avx}(V)$)
\item  idéal engendré par $\seq{r}{m} \in \avx$ dans $\avx$:
$$<\seq{r}{m}>_{\avx}:=\{u_1r_1+\cdots + u_mr_m \mid \seq{u}{m} \in \avx\}$$
(par défaut $<\seq{r}{m}>:=<\seq{r}{m}>_{\avx}$)
\item $r \in \avx$ : idéal monogène $r\avx :=<r>_{\avx}$ 
\item $I$ {\it maximal} dans $\avx$ si $(\forall r \in \avx - I) \; I + <r>=\avx$ 
\item $I_1,I_2$ {\it comaximaux} si $I_1+I_2 = \avx$
\item $I$ {\it radical} si $r^m \in I \Rightarrow r \in I$ 
\item $I,J$ idéaux de $\avx$ ; {\it injecteur de $I$ dans $J$} :
$$\Inj({I},{J}):=\{\sigma \in \SY_n \mid \; \sigma.I \subset J\}$$
\item si $V:=V(I)$ est finie alors 
\begin{eqnarray}
\dim_k(\avx/{\Id(V)}) = \Card(V) 
\end{eqnarray}
\item si $V(I)$ est finie alors $\avx/I$ est engendré par les monômes sous l'escalier d'un base de Gr\"obner de $I$ pour l'ordre lexicographique.
\end{itemize}

{\bf 4.2 Idéaux galoisiens : notations et définitions }\\

\begin{itemize}
\item tout idéal galoisien est radical
\item $H \subset \SY_n$ , l'{\it idéal galoisien $I_\racf^H$ défini par $H$ et $\racf$} :
\begin{eqnarray*}
I_\racf^H & : = &\{r \in \avx \mid (\forall \sigma \in H) \; \sigma.r(\racf) = 0 \}\\
\end{eqnarray*}
\item $I_\racf:=I_\racf^{I_n}$ 
\item {\it idéal des $\racf$-relations} : $\MFM := \{r \in \avx \mid r(\racf)=0\}=I_\racf$
\item {\it idéal des relations symétriques} : $\MFS:=I_\racf^{\SY_n}$ ; ne dépend pas du choix de $\racf$.
\item Autres façons de voir l'idéal galoisien $I_\racf^H$ définit par $H$ et $\racf$ :
\begin{eqnarray*}
I_\racf^H & = &\{r \in \avx \mid (\forall \sigma \in H) \; r(\sigma*\racf) = 0 \}\\
& = & \Id(H*\racf)\\
&=& \{r \in \avx \mid (\forall \sigma \in H) \; \sigma.r\subset \MFM \}\\\end{eqnarray*}
(on pourra noter $I_\MFM^H:=I_\racf^H $ ).
\item Attention : si $I_n \not \subset H$ alors $I_\racf^H \not \subset \MFM=I_\racf$
\end{itemize}

%\section{}
{\bf 4.3 Ensembles de permutations particuliers} % Idéaux et ensembles de permutations
% {\bf Définitions et notations}
\begin{itemize}
\item {\it groupe de décomposition de $I$} (aussi Stab$_{S_{n}}(I)$ ) : $Gr(I)=\Inj({I},{I})$
\item $\Max({I},{\racf})$ : {\it plus grand ensemble de permutations définissant $I_\racf^H$ avec $\racf$}
\item {\it groupe de Galois de $\racf$ sur $k$} : 
\begin{eqnarray}
Gal_k(\racf):=Gr(\MFM)&=&  \Max({\MFM},{\racf})\\
&=&\{\sigma \in \SY_n \mid \; r(\racf)=0 \Rightarrow \sigma.r(\racf)=0\}
\end{eqnarray}
\end{itemize}

{\bf 4.4 Idéaux galoisiens purs}

$I$ est dit {\it pur} si $\Max(I,\racf)$ est un groupe. 

Nous retrouverons ces idéaux galoisiens plus loin.

{\bf 4.5 Premières propriétés}

$I,J,I_i$ idéaux galoisiens, $G,H \subset \SY_n$, $\sigma,\tau \in \SY_n$, $\MFM=I_\racf$

\begin{itemize}
\item $\MFM$ est  un idéal maximal de $\avx$\\
($r(\racf)=0 \Rightarrow (\forall \sigma \in \Gal_k(\racf)), 
\sigma.r(\racf)=0 \Rightarrow r \in \MFM$)
%\item  ${\mathcal I}$ est un idéal galoisien si et seulement si ${\mathcal I}$ idéal t.q.
 %$\MFS \subset {\mathcal I} \neq \avx$
\item Cette propriété et la suivante sont exprimées sous une autre forme dans 4.6
         $$I_{\sigma * \racf}^H = I_\racf^{\sigma H}$$
\item $$\sigma.I_\racf^{H}=I_\racf^{H\sigma{-1}}$$ 
\item C.P. $I_{\sigma * \racf}=\sigma^{-1}.I_\racf=\sigma^{-1}.\MFM$ 
\item ${\mathcal G}$ 
un ensemble d'ensembles de permutations (i.e. inclus dans les parties de $\SY_n$ ) ;
$$
I_\racf^{\bigcup_{G \in {\mathcal G}}} = \bigcap_{G \in {\mathcal G}} I_\racf^G
$$
\item %Soit $I=I_\racf^H$
\begin{eqnarray}
 \Max({I},{\racf})&=&\Inj({I},{\MFM})
=\{\sigma \in \SY_n \mid \; r \in I \Rightarrow \sigma.r(\racf)=0\}\\
&=& \Gal_k(\racf)H\\
&=& \bigcup_{G \subset \SY_n \mid I=I_\racf^G} G
\end{eqnarray}
\item $$ I \subset I_\racf ^{\Gr(I)} \quad (\forall \racf \in V(I)) $$

\item si $H$ est un groupe alors
$$
I_\racf^H=I_\racf^{\Gr(I_\racf^H)} 
$$
et $$
H \subset \Gr(I_\racf^H) \subset Max(I_\racf^H,\racf) 
$$

\item $\Max(I,\racf)$ est un groupe (i.e. $I$ est pur) si et seulement si l'une des conditions suivantes est satisfaite 
\begin{enumerate}
 \item[(i)] $\Gal_k(\racf) < Gr(I)$
\item[(ii)] $Gr(I)=\Max(I,\racf) (=\Inj(J,\MFM))$
\item[(iii)] $Gr(I)=\Inj({I},{J})$ pour tout idéal $J$ contenant $I$
%\item[(iv)] 
%\item (iii) $Gr(I) 
\end{enumerate} 
\item $(\forall \beta \in V(I))$, $\sigma \not \in \Max(I,\uplet{\beta})$ si et seulement si $$I +\sigma.I = \avx$$
(équivalent à : pour tout idéal maximal $\MFM^\prime$ contenant $I$,  $\sigma \not \in \Inj(I,\MFM^\prime$))
\item $\SY_n=Gr(\MFS)=\Max({\MFS},{\racf})$ 
\item  {\bf Correspondance galoisienne inhérente aux idéaux galoisiens}
\begin{enumerate}
\item[(i)] $G \subset H \subset \SY_n \Rightarrow I_\racf^H \subset I_\racf ^G$
\item[(ii)] $\MFS \subset {\mathcal I}\subset {\mathcal J}\subset \MFM=I_\racf \Rightarrow {\mathcal I},{\mathcal J}$ sont galoisiens et 
 $$\SY_n\subset \Max({{\mathcal I}},{\racf})\subset \Max({{\mathcal J}},{\racf}) \subset \Grac$$
\end{enumerate}
\item Comme $\MFM=I_\racf^{I_n}=I_\racf^\Grac$ et $\Grac =Gr(\MFM)=\Max(\MFM,\racf)$, $$(\forall H \subset \Grac), \quad I_\racf^H=I_\racf^\Grac$$
\end{itemize}
%{\ }\\
{\bf 4.5 Autres façon d'exprimer $I\subset I_\racf=\MFM$}
\begin{itemize}
\item 
\begin{eqnarray*}
I& = & \bigcap_{\sigma \in \Max(I,\racf)} I_{\sigma*\racf}\quad \\
&= & \bigcap_{\sigma \in \Inj(I,\MFM)}\sigma^{-1}.\MFM
\end{eqnarray*}
\item %Montrer l'égalité 
$\MFS \subset I \subset J \Rightarrow \bigcap_{\sigma \in \Inj(I,J)}\sigma^{-1}.J \subset I$
\item $\MFS \subset I\subset J \subset \MFM$,
$H=\Max({I},{\racf})$, $G=\Max({J},{\racf})$ t.q.
$$H=G\tau_1+\ldots +G\tau_s \qquad (*)$$
alors
\begin{eqnarray}
 I = \bigcap_{i=1}^s\tau_i^{-1}.J 
  \end{eqnarray}   
(on peut montrer que la condition $(*)$ est toujours satisfaite)
\item C.P. $G=\Gal_k(\racf)$. Soient $\tau_1,\ldots ,\tau_s$ t.q. 
$\Max(I,\MFM)= G\tau_1+ \ldots +G\tau_s$ alors
\begin{eqnarray}
 I &= &\bigcap_{i=1}^s\tau_i^{-1}.\MFM \quad (=\bigcap_{i=1}^sI_{\tau_i.\racf})
 \end{eqnarray}   
\item C.P. Soit $\tau_1=id,\ldots ,\tau_s$ transversale à droite de $\SY_n$ modulo $\Gal_k(\racf)$.
\begin{eqnarray}
\MFS &= &\MFM \cap \tau_2^{-1}.\MFM \cap \cdots \cap \tau_s^{-1}.\MFM \quad 
          ( =  I_\racf \cap I_{\tau_2.\racf}\cap \cdots \cap I_{\tau_s.\racf} )
    \end{eqnarray}      
\item Soit ${\mathcal E}$ un ensemble d'idéaux galoisiens
 $$ I = \bigcap_{J \in {\mathcal E}} J \Rightarrow \Gr(I) =\bigcap_{J \in {\mathcal E}} \Inj(I,J) $$
 \end{itemize}

{\bf 4.6 Quelques propriétés des injecteurs}
\begin{itemize}
\item \begin{eqnarray}\Inj({\sigma.I},{\tau.J})=\tau\Inj({I},{J})\sigma^{-1} \end{eqnarray}
\item C.P. $\Gal_k(\sigma.\racf)=\Inj({\sigma^{-1}.\MFM},{\sigma^{-1}.\MFM})
                =\sigma^{-1}\Gal_k(\racf)\sigma$
 \item $H\Inj(I,\MFM)=\Inj(.I,\MFM) \Rightarrow H\Inj(G.I,\MFM) = \Inj(G.I,\MFM) $               
 \item   $\MFS \subset I_i   \subset J$, $i=1,2$ ;
 \begin{eqnarray}
   \Inj({I_1 + I_2},{J} )        = \Inj({I_1},{J}) \cap \Inj({I_2},{J})
\end{eqnarray}   
\item $\MFS \subset I  \subset J_i$, $i=1,2$  ;% preuve : feuille verte 
$\Inj(I,J_1\cap J_2) = \Inj(I,J_1)\cap \Inj(I,J_2)$
   \item $H \subset \SY_n$, $L:= \Inj({I},{J})$ ; 
   $$\Inj({H.I},{J} )= \bigcap_{h \in H}Lh^{-1} \text{ et } \Inj({I},{H.J}) = \bigcap_{h \in H} hL$$
\item % Montrer l'égalité inclusion facile
$ \Inj(J,\MFM)\Inj(I,J) \subset \Inj(I,\MFM)$
\item Soit un polynôme $f$ de degré $n$ et $I$ un idéal galoisien de $f$ d'injecteur un groupe $H$. L'ensemble des idéaux galoisiens de $f$ d'injecteur $H$ est formé des $\sigma . I$ où $\sigma$ parcourt le normalisateur de $H$ dans $S_{n}$. 
\end{itemize}

\section{Variétés}
{\bf 5.1 Généralités classiques}

$I,I_1,I_2$ idéaux de $\avx$
\begin{itemize}
\item variété de $I$ : $V(I) := \{\uplet{\beta} \in \ck^n \mid \; (\forall r \in I) \; r(\uplet{\beta})=0\}$
\item $V(\Id(V(I)))=V(I)$
\item $I = I_1 \cap I_2 \Rightarrow V(I)=V(I_1) \cup V(I_2)$
\item $I_1, I_2$ comaximaux $\Leftrightarrow$ $V(I_1) \cap V(I_2)=\{\ \}$
\item $I$ radical $\Leftrightarrow I=\Id(V(I))$
\item $I_1,I_2$ radicaux ; $V(I)=V(I_1) \cup V(I_2) \Rightarrow I = I_1 \cap I_2$

\end{itemize}

{\bf 5.2 Idéaux galoisiens}

$I$ idéal galoisien
 %\subset \MFM=I_\racf$
 
\begin{itemize}
\item \begin{eqnarray}
V(I) = Max(I,\racf)\star\racf
\end{eqnarray}
\item C.P.  $V(\MFS)=\SY_n\star \racf \quad {et } \quad V(\MFM)=Gal_k(\racf)\star \racf$
%\item $V(\Id(H\star \racf))=H\star \racf$ si et seulement si $H=Max(I,\racf)$.
\end{itemize}
La variété d'un idéal galoisien peut s'exprimer trivialement sous forme de variétés disjointes dès lors que l'on exprime $Max(I,\racf)$ sous forme de classes à droite disjointes $G_{\racf}\tau$ et que $f$ est sans racine multiple.

\section{Anneaux quotients, Variétes, Idéaux galoisiens et Injecteurs}

Ce paragraphe doit être considéré avec attention pour concevoir qu'un même résultat peut s'exprimer de nombreuses fa{\c c}ons selon qu'on l'exprime avec les variétés, les injecteurs, les anneaux quotients ou les idéaux galoisiens et/ou en modifiant légèrement les hypothèses pour lui donner une apparence de nouveauté.

$I=I_{\racf}^L$, on a $Max(I,\racf)=G_{\racf}\tau_{1}+\cdots +G_{\racf}\tau_{s}$, $\tau_{i} \in L$ (on peut toujours supposer que
$L=Max(I,\racf)$). Un idéal galoisien d'un polynôme sans racine multiple est radical. Notre polynôme $f$ est supposé sans racine multiple.

\begin{itemize}
\item  $$\dim_k(\avx/{I}) = \Card(V(I)) = \Card(Max(I,\racf))$$
\item C.P. idéal de $\racf$-relations :
$$ \avx/{\MFM} \equiv k(\racf)$$
et
$$\dim_k(\avx/{\MFM}) =  \Card(\Gal_k(\racf))$$
\item C.P. idéal des relations symétriques
$$\dim_k(\avx/{\MFS}) = n!$$
\end{itemize}

Soient $V_{1},\ldots, V_{s}$  des variétés incluses dans $V(I)$  et $L_{1},\ldots ,L_{s}$ les ensembles maximaux de permutations tels que $V_{i}=L_{i}.\racf$ (i.e. $L_{i}=G_{\racf}=Max(Id(V_{i}),\racf)$).

L'ensemble $V(I)$ est l'union disjointe des variétés $V_{1},\ldots, V_{s}$ ssi  les ensembles de permutations $L_{1},\ldots ,L_{s}$  sont deux-à-deux disjoints et (union disjointe) 
$$Max(I,\racf)=L_{1}+\cdots + L_{s} ;$$ 
dans ce cas, puisque les $I_{i}:=Id(V_{i})$ sont deux-à-deux comaximaux et que $I = \bigcap_{i=1}^sI_{i}$, il vient (théorème des restes chinois) :

$$
\avx/{I} = \prod_{i=1}^s \avx/{I_{i}} 
$$

En particulier, soient $\tau_{1},\ldots ,\tau_{s}$ (deux-à-deux distincts)  tels que (union disjointe)
$$Max(I,\racf)=G_{\racf}\tau_{1}+\cdots + G_{\racf}\tau_{s} ;$$ 
comme $I_{i}=\tau_{i}^{-1}.\MFM$ en posant $I_{i}=\MFM_{i}$, un idéal maximal conjugué de $\MFM$,
$$
\avx/{I} = \prod_{i=1}^s \avx/{\MFM_{i}} \equiv  k(\racf)^s
$$

\section{Polynômes multivariés particuliers}
\begin{itemize}
\item $e_0(\vary)=1,e_1(\vary), \ldots ,e_r(\vary),\ldots$, les {\it fonctions symétriques élémentaires} en $\vary:=(y_1,\ldots ,y_n)$ : 
\begin{eqnarray*}
e_r(\vary) :=& \sum_{m \in \SY_n.(y_1\cdots y_r)} m &\quad \text{si $1\leq r \leq n$}\\
e_r(\vary) :=& 0 &\quad \text{si $ n < r$}
\end{eqnarray*}
$e_1(\vary)=y_1+\cdots + y_n$, $e_2(\vary)=y_1y_2 + y_1y_3+\cdots + y_{n-1}y_n$, 
\ldots , $e_n(\vary)=y_1y_2\cdots y_n$.
\item $p_0(\vary)=n,p_1(\vary), \ldots ,p_r(\vary),\ldots$, les {\it fonctions puissances} (de Newton) en $\vary:=(y_1,\ldots ,y_n)$ : 
\begin{eqnarray*}
p_r(\vary) :=& \sum_{i=1}^n y_i^r &\quad (\forall r \in \N)\\
\end{eqnarray*}
\item $h_0(\vary)=1,h_1(\vary), \ldots ,h_r(\vary),\ldots$, les {\it fonctions complètes} en $\vary:=(y_1,\ldots ,y_n)$ : 
\begin{eqnarray*}
h_r(\vary) :=& \sum_{i_1+\cdots +i_n=r} \vary^{\uplet{i}} &\quad (\forall r \in \N)\\
\end{eqnarray*}
\item $e_r:=e_r(\varx)$, $p_r:=p_r(\varx)$,\ldots ,$\widetilde{e_r}=e(\racf)$,\ldots
\item {\it formules de Girard-Newton} : pour tout $m \geq 0$
\begin{equation}
\label{eq:formuleGirardNewton}
p_me_0 - p_{m-1}e_1 + \cdots + (-1)^{m-1}p_1e_{m-1}+ (-1)^mm.e_m = 0 \quad .
\end{equation} 
\item {\bf Polynôme $f=a_nx^n+a_{n-1}x^{n-1}+\cdots + a_0$ : }
 $$\frac{f}{a_n}= x^n -e_1(\racf)x^{n-1}+e_2(\racf)x^{n-2}+\cdots + (-1)^ne_n(\racf)$$
 \item {\it formules de Girard-Newton} : $f=a_nx^n+a_{n-1}x^{n-1}+\cdots + a_0$ ; pour tout $m \geq 0$
\begin{equation}
%i.e. $a_{n-i} = (-1)^{e_i}$
\label{eq:formuleGirardNewton}
\widetilde{p_m}a_0 + \widetilde{p_{m-1}}a_1 + \cdots + \widetilde{p_1}a_{m-1}+ m.a_m = 0 \quad .
\end{equation} 
où $a_i=0$ si $i> n$.
\item $C_1(f),\ldots ,C_n(f)$, {\it les modules de Cauchy de $f$} dans $\avx$ :
posons $C_i:=C_i(f)$
\begin{eqnarray*}
C_1(x_1) &:=& f(x_1) \\
C_{2}(x_1,x_2) &:=& \frac{ C_1(x_{1})-C_1(x_2)}{x_{1}-x_2}\\
%C_{3}(x_{1},x_{2},x_3) &=& \frac{C_2(x_1) - C_1(x_2)}{x_1-x_2}\\
&\vdots &\\
C_{r}(x_{1},\ldots ,x_{r}) &:=& \frac{C_{r-1}(x_{1},\ldots , x_{r-2}, ,x_{r-1})-C_{r-1}(x_{1},\ldots ,x_{r-2}, ,x_{r}) }{x_{r-1}-x_{r}} \, \quad 1<r\leq n \\
\end{eqnarray*}

%\begin{eqnarray*}
%C_n(x_n) &:=& f(x_n) \\
%C_{n-1}(x_{n-1},x_n) &:=& \frac{ C_n(x_{n-1})-C_n(x_n)}{x_{n-1}-x_n}\\
%C_{n-2}(x_{n-2},x_{n-1},x_n) &=& \frac{C_2(x_1) - C_1(x_2)}{x_1-x_2}\\
%&\vdots &\\
%C_{r}(x_{r},\ldots ,x_{n}) &:=& \frac{C_{r+1}(x_{r},x_{r+2},\ldots ,x_{n})-C_{r+1}(x_{r+1},x_{r+2},\ldots ,x_{n})}{x_{r}-x_{r+1}} \, \quad 1<r\leq n \\
%\end{eqnarray*}

\item 
\begin{eqnarray}\label{eq : Cauchy}
C_{r+1}=\sum_{i=r}^{n}a_i h_{i-r}(x_1,x_2,\ldots ,x_{r}) \quad r=0,\ldots ,n -1
%C_{r}=\sum_{i=r}^{n}a_ih_{i-(n-r)}(x_{r},x_{r+1},\ldots ,x_{n}) \quad r=1,\ldots ,n \quad 
\end{eqnarray}
\item EX. $f:=x^4 - 2x^3 + 2x^2 + 2$
\begin{eqnarray*} 
C_1(x_1)&=&x_1^4-2x_1^3+2x_1^2+2\\
C_2(x_1,x_2)&=&x_2^3+x_1x_2^2+x_1^2x_2 +x_1^3 -2(x_2^2+x_1x_2+x_1^2)+2(x_2+x_1)\\
C_3(x_1,x_2,x_3)&=&x_3^2+x_3x_2+x_2^2+ x_3x_1+x_2x_1+x_1^2 -2(x_2+x_1+x_3)+2\\
C_4(x_1,x_2,x_3,x_4)&=&x_4+x_3+x_2+x_1-2 \quad .
\end{eqnarray*}
\end{itemize}

\section{Polynômes caractéristiques, minimaux et résolvantes}

{\bf Généralités classiques}\\

$\Theta \in \avx$, $I$ idéal de $\avx$ 
\begin{itemize}
\item {\it endomorphisme multiplicatif induit} par $\Theta$ dans $\avx/I$ :
 \begin{eqnarray*} \hat{\Theta} :
      \avx/I \longrightarrow& \avx/I  , \quad \overline{p}  \mapsto  \hat{\Theta}(\overline{ p})=\overline{\Theta.p}
       \end{eqnarray*}   
\item $C_{\Theta,I}$, $M_{\Theta,I}$ : 
polynômes caractéristique et minimal de $\hat{\Theta}$
\item si $k$ parfait alors $M_{\Theta,I}$ est la forme sans facteur carré de $C_{\Theta,I}$
\item si $I$ radical alors
\begin{eqnarray}
C_{\Theta,I} = \prod_{\uplet{\beta} \in V(I)} (x- \Theta(\uplet{\beta}))
\end{eqnarray}
\end{itemize}

{\bf Hypothèses}\\

$H\subset L \subset \SY_n$, $I=I_\racf^L$, $L=Max(I,\racf)$, $\Theta \in \avx$, $H=\Stab_L(\Theta)$.

{\bf Invariants}
\begin{itemize}
\item $\Theta$ est un {\it $H$-invariant $L$-primitif } si $H=\Stab_L(\Theta)$\\
on pourra lire relatif à la place de primitif 
\item  Soit $K=k(e_{1},\ldots ,e_{n})$. $\Theta$ est l'élément un {\it $H$-invariant $L$-primitif } ssi $\Theta$  est un élément $K$-primitif de $K(\gx)^H$ sur $K(\gx)^L$
\item dans MAGMA : {\tt RelativeInvariant} dans GAP : {\tt PrimitiveInvariant}
\item $\Theta$ est un $H$-invariant $L$-primitif {\it $\racf$-séparable} si pour tout $\sigma \in L$
% l'une des conditions équivalentes suivantes est satisfaite :
%\begin{enumerate}
%\item[(i)] $R$ est sans racine multiple
%\item[(ii)] 
$$\sigma.\Theta \neq \sigma.\Theta  
               \Rightarrow \sigma.\Theta(\racf) \neq \tau.\Theta(\racf)$$

\item soit $M$ un sous-groupe de $L$ ; alors tout $H$-invariant $L$-primitif  est aussi un $H$-invariant $M$-primitif 
\item tout $H$-invariant $L$-primitif $\racf$-séparable est aussi un $H$-invariant $M$-primitif $\racf$-séparable
\item $f$ sans racine multiple ; il existe toujours un $H$-invariant $\SY_n$-primitif $\racf$-séparable pour tout sous-groupe $H$ de $\SY_n$
\item un $H$-invariant $L$-primitif est dit {\it universel} s'il est toujours $\racf$-séparable pour tout $n$-uplet $\racf$ constitué de valeurs deux-à-deux distinctes 
\item EX. le Vandermonde $\prod_{1\leq i < j \leq n}(x_i-x_j)$ est un $A_n$-invariant $\SY_n$-primitif universel.
\item $\uplet{i} \in \N^n$ ; un $H$-invariant (on peut prendre aussi une combinaison linéaire) 
:
$$
\Psi_{\uplet{i} ,H}:=\sum_{\sigma \in H} (\sigma \star \varx)^{\uplet{i}}
$$
\item si les parts de $\uplet{i}$ sont deux-à-deux distinctes alors $\Psi_{\uplet{i} ,H}$ est $\SY_n$-primitif 
(car $\sigma . (H . \varx^{\uplet{i}})=H . \varx^{\uplet{i}} \Leftrightarrow \sigma \star (H \star \varx)=H \star \varx \Leftrightarrow \sigma H =H$)
\item $\sigma \in L$ ; si $\Theta$ $H$-invariant $L$-primitif alors $\sigma.\Theta$ est un
$H^\sigma$-invariant $L$-primitif
\end{itemize}

{\bf Résolvantes}\\

\begin{itemize}
\item $G\subset \SY_n$, $\Theta \in \avx$, {\it résolvante $G$-relative de $\racf$ par $\Theta$} :
$$R_{\Theta,G,\racf}:=\prod_{\Psi \in G.\Theta}(x-\Psi(\racf))$$
(on peut prendre $\Theta \in \cvx$ si aucun dénominateur des polynômes de $G.\Theta$ n'appartient à $\MFM$ ; i.e. il ne s'annule pas en $\racf$)
\item on peut noter $R_{\Theta,I}:= R_{\Theta,\Max(I,\racf),\racf}$ 

\item une résolvante $L$-relative de $\racf$ par $\Theta$ est appelée une {\it $H$-résolvante $L$-relative de $\racf$}
\item $$C_{\Theta,I}=R_{\Theta,I}^{\#H}$$
\item si $k$ corps parfait alors $R_{\Theta,I} \in \ax$
\item si $L=\tau_1H+\cdots + \tau_eH$ alors
$$
R_{\Theta,I}=\prod_{i=1}^e(x-\tau_i.\Theta(\racf))
$$
\item $\Theta$ est un $H$-invariant $L$-primitif {\it $\racf$-séparable} si et seulement si
$R_{\Theta,I}$ est sans racine multiple (i.e. $i\neq j \Rightarrow \tau_i.\Theta(\racf) \neq \tau_j.\Theta(\racf)$) ; dans ce cas
$R_{\Theta,I}$ est la forme sans facteur carrée de $C_{\Theta,I}$, le polynôme caractéristique de $\hat \Theta$, et si, de plus, $k$ est parfait alors $R_{\Theta,I}=M_{\Theta,I}$, le polynôme minimal de $\hat \Theta$.

\item Soit $K=k(e_{1},\ldots ,e_{n})$. La résolvante générique $R_{\Theta,L,\gx}$ est le polynôme minimal de $\Theta$ sur $K(\gx)^L$ 
\item Tout facteur $h$ $k$-irréductible de $R_{\Theta,I}$ est naturellement le polynôme minimal de chacune de ses racines sur $k$ ; \item ($k$ parfait) si $h$ est un facteur $k$-irréductible de multiplicité 1 de la résolvante $R_{\Theta,I}$  et ne possédant aucune racine multiple alors chacune de ses racines $\sigma.\Theta(\racf)$ est un élément
primitif de $k(\racf)^{G_{\racf} \cap H^\sigma}$ sur $k(\racf)^{G_{\racf} \cap L}$ où $G_{\racf}$ est le groupe de Galois de $\racf$ sur $k$ (ici $k(\racf)^{G_{\racf} \cap L}=k(\racf)^{G_{\racf}}=k$ puisque $G_{\racf} \subset L$, par hypothèse.
\item $h$ est un facteur simple sur $k$ de la résolvante $R_{\Theta,I}$ ssi $\Theta$ est un $H$-invariant $G_{\racf}$-primitif $\racf$-séparable
\item Pour toutes les subtilités concernant les facteurs des résolvantes (simples, non simples, $k$ parfait ou non parfait) voir
\cite{VALIBOUZE:2009:HAL-00421725:1}
\item pour étudier les résolvantes avec les classes doubles : voir \cite{Valibouze:05}
%\end{enumerate}
\end{itemize}
\section{Générateurs des idéaux galoisiens}
{\bf Hypothèses :}\\
%$\MFS \subset I \subset \MFM$, 
$I$ idéal galoisien, $\racf \in V(I)$, $f$ sans racine multiple, $k$ corps parfait (si $g \in \ax$ est irréductible alors $g$ est sans racine multiple)

\begin{itemize}
\item
\begin{eqnarray}
\MFS &= &<e_1-\widetilde{e_1},\ldots , e_1-\widetilde{e_1}>\\
          &=& <C_1,\ldots ,C_1>
          \end{eqnarray}
          
\item {\bf Théorème de l'élément primitif sur les idéaux galoisiens} \\

Un polynôme $R$ est dit {\it $J$-primitif de l'idéal $I$} si $J = I +<R>$.

$I$ idéal galoisien, $L=Max(I,\racf)$,  $H\subset L$, $\Theta$ un $H$-invariant $L$-primitif $\racf$-séparable alors 

\begin{eqnarray}
I_\racf^H = I + < h(\Theta)>
\end{eqnarray}

où $h$ est le facteur simple de $R_{\Theta,I}$ tel que $h(\Theta(\racf))=0$

\item (version effective) Soit $h$ un facteur simple de $R_{\Theta,I}$ et $\uplet{\beta}$ tel que 
$h(\Theta(\uplet{\beta}))=0$ alors

\begin{eqnarray}
I_{\uplet{\beta}}^H = I + < h(\Theta)>
\end{eqnarray}

\item {\bf Racines multiples de la résolvante} : le théorème de l'élément primitif et l'usage des matrices de groupes se généralisent. Consulter l'article \cite{VALIBOUZE:2009:HAL-00421725:1} dans lequel un exemple concret conclut à partir d'un facteur multiple de la résolvante.

Ci-après une partie des résultats relève de l'article \cite{AubryValibouze:00}.

\item $\{f_1,\ldots ,f_n\} \in \avx $ est un {\it ensemble triangulaire} si pour tout $i\in \intervalle{1}{n}$, le polynôme $f_i$ est de degré au minimum 1 en $x_i$ et s'exprime sous la forme :
$$
f_i = x_i^{deg_{x_i}(f_i)} + g(x_1,\ldots ,x_i) 
$$
%où $m_i = >0$ 
\item un ensemble triangulaire $\{f_1,\ldots ,f_n\}$ est dit {\it séparable} si $f_1$ est sans racine multiple pour tout $i\in \intervalle{2}{n}$ et pour tout $\uplet{\beta} \in V(<f_1,\ldots ,f_{i-1}>)$, $f_i(\beta_1,\ldots ,\beta_{i-1},x)$ est sans racine multiple.
\item
si $I$ est pur et  $f$ sans racine multiple alors il existe un ensemble triangulaire séparable engendrant $I$ ; cet ensemble forme une base de Gr\"obner de $I$ pour l'ordre lexicographique :
$$
I=<f_1,\ldots , f_n > ;
$$
\item $\Init(I):=(\deg_{x_1}(f_1),\ldots , \deg_{x_n}(f_n))$ : $n$-uplet constitué des {\it degrés initiaux} de $I$
\item $H\subset \SY_n$, $H_{(0)} := H$, $H_{(i)}:=Stab_{H_{(i-1)}}(i)$, $i=1,\ldots n$
\item  $\Gal_k(\racf) < H=\Gr(I)$ (i.e. $I$ pur), $(\forall i\in \intervalle{1}{n})$,  

$$ (\forall \uplet{\beta} \in V(I)) \quad f_i(\beta_1,\ldots ,\beta_{i-1},x) = \prod_{\tau \in H_{(i-1)}/H_{(i)}}(x -\beta_{\tau(i)})$$
(généralisable à tout idéal galoisien triangulaire)

\item  $\Gal_k(\racf) < H=\Gr(I)$ (i.e. $I$ pur), pour tout $i\in \intervalle{1}{n}$ ; alors
$$\deg_{x_i}(f_i) = \frac{\#H_{(i-1)}}{\#H_{(i)}}$$
%\item $m_i:=\frac{\#H_{i-1}}{\#H_{i}}$
\item $I$ triangulaire, $\Init(I)=(m_1,\cdots ,m_n)$, degrés initiaux ;
$$\avx/I =<\varx^{\uplet{i}} \mid \; 0\leq  i_1 < m_1,\ldots 0 \leq , i_n < m_n >$$
\item $I$ galoisien triangulaire ;
\begin{eqnarray}
\dim_k(\avx/I) &=& \Card(V(I)) \\
&= &\Card(\Max(I,\racf)) \nonumber \\
&=&\prod_{i=1}^n \deg_{x_i}(f_i) \nonumber
\end{eqnarray}
\item si $I=\MFM=<f_{1},\ldots ,f_{n}>$ ($\MFM$ est pur) où les $f_{i}$ sont appelés les {\it modules fondamentaux} par Tchebotarev. \item Chaque module fondamental $f_{i}$ est le polynôme minimal de $\alpha_{i}$ sur $k(\alpha_{1},\ldots ,\alpha_{i-1})$. 

\end{itemize}

L'objectif de la théorie de Galois effective (et donc du projet Galois) est de calculer  l'idéal des relations $\MFM$ et le groupe de Galois $\Gr(\MFM)$, son groupe de décomposition (un groupe de décomposition est facilement calculable), c'est-à-dire le corps de décomposition de $f$ avec l'action du groupe de Galois sur ses racines.

Pour ce faire, il faut utiliser toutes les méthodes pour calculer des facteurs dans les extensions ou bien calculer une base de 
Gr\"obner d'un idéal maximal :
incluant et combinant de l'interpolation multivariée (voir \cite{Lederer:2004}, \cite{McKayStauduhar:97}) avec du numérique, du modulaire, ...), de l'algèbre linéaire (avec du numérique, du $p$-adique : \cite{Yokoyama:99} et \cite{Yokoyama:97}, du modulaire), des résolvantes, des matrices de groupes, la construction d'une chaîne ascendante d'idéaux galoisiens, ...; on peut supposer connaître un idéal $I$ (à défaut l'idéal des relations symétriques) et chercher un seul idéal de sa décomposition en idéaux premiers (ils sont tous conjugués) en utilisant les injecteurs pour accélérer et orienter les calculs (voir \cite{Valibouze:99}). Les informations, même partielles, sur le groupe de Galois sont exploitables pour éviter de nombreuses étapes de calculs, voir pour calculer directement $\MFM$.

Et pour tout dire, il faut mélanger toutes les méthodes pour produire un algorithme parallèle et collaboratif. 

Il convient néanmoins de concevoir que les stratégies sont distinctes des techniques de calcul. Par exemple, l'interpolation multivariée peut se réaliser en modulaire ou en numérique. L'interpolation multivariée étant elle-même une technique de calcul reposant sur des informations galoisiennes. L'objectif d'établir une formule exprimant les ``séparateurs'' (i.e. les polynômes jouant le rôle des polynômes de Lagrange en univarié) dans le cas d'un idéal galoisien connaissant son injecteur ; dans le cas $I=\MFM$ pour \cite{Lederer:2004} restreint aux modules fondamentaux  $f_{i}(\alpha_{1},\ldots,\alpha_{i-1},x)$ linéaires en $x$ pour \cite{McKayStauduhar:97}.

Nous avons désormais des outils simples et extrêmement efficaces : par exemple, permuter une relation peut produire un module fondamental ou un élément primitif d'un idéal galoisien ; cette permutation est prévisible avant l'exécution du programme (voir Section sur les idéaux galoisiens purs ou bien \cite{Valibouze:08}, \cite{Valibouze:95} et \cite{Sargov:expmath}). Ce qui signifie que les calculs sont dans ces cas instantanés. Un autre cas de calculs rapides et prévisibles : il suffit de diviser certains générateurs de $\MFM$ par d'autres pour obtenir immédiatement de nouveaux. Les exemples donnés dans  \cite{Valibouze:08} illustrent l'efficacité de ces deux techniques. Cette méthode inclut celle consistant à calculer des modules de Cauchy de certains $f_{i}$ pour en déduire d'autres (voir  \cite{Sargov:expmath}).
Il est facile de reformuler ces méthodes en modifiant le contexte sans rien apporter de nouveau pour le calcul effectif.

\section{Idéaux galoisiens purs}
Rappelons que $I$ est dit {\it pur} si $\Max(I,\racf)$ est un groupe.

Par définition : $Gr(I)=Stab_{S_{n}}(I)=\Inj(I,I)$ et $\Max(I,\racf)=\Inj(I,\MFM)$.

On peut toujours se ramener à un idéal galoisien pur lorsque les constructions d'idéaux galoisiens sont réalisées à partir de groupes (ce qui est le cas en pratique) (voir \cite{Valibouze:05}) :

Si $H$ et $E$ sont deux sous-groupes de $\SY_n$ tels que 
$\Inj(I,\MFM)=HE$
et que $H$ est un tel groupe qui est maximal dans $\Inj(I,\MFM)$ alors $H.I$ est un idéal galoisien pur de groupe de décomposition $H$. 

Le groupe $H$ existe toujours puisque si $I=I_{\racf}^E$ alors $\Inj(I,\MFM)=G_{\racf}E$, $G_{\racf}$, groupe de Galois de $\racf$ sur $k$.

Le choix du nombre minimal de permutations de $H$  suffisantes au calcul de $H.I$ (en effectif, les permutations portent sur les générateurs de $I$) est étudié dans  \cite{Valibouze:08} et \cite{Sargov:expmath} offre une étude pratique en degré 8. Prévoir ces permutations est pré-calculable groupistiquement (c'est ce qui a été fait pour l'étude de cas dans \cite{Sargov:expmath}).

Si un idéal galoisien est pur alors il est triangulaire (voir \cite{AubryValibouze:00}).

Nous avons dans ce qui précède de nombreuses conditions nécessaires et suffisantes pour qu'un idéal galoisien soit pur.
Nous les récapitulons ici. Les conditions suivantes sont équivalentes 
\begin{enumerate}
\item $I$ est pur
\item $\Max(I,\racf)$ est un groupe
 \item $\Gal_k(\racf) < Gr(I)$ 
\item $Gr(I)=\Max(I,\racf)$ 
\item $Gr(I)=\Inj({I},{J})$ pour tout idéal $J$ contenant $I$ (si c'est vrai pour $\MFM$, c'est vrai pour tous)
\item $$\dim_k(\avx/{I}) = \Card(\Stab_{S_{n}}(I))$$
\item $$\Card(V(I)) = \Card(\Stab_{S_{n}}(I))$$
\item $$ \prod_{i=1}^n \deg_{x_i}(f_i)=\Card(\Stab_{S_{n}}(I))$$
\end{enumerate}

%$C_1(\seq{x}):= x_1+\cdots +x_n -\frac{a_{n-1}}{a_n}$

\section{Corps des racines}

{\bf Théorie (presque)-classique}

{\bf Généralités classiques : rappels}

\begin{itemize}
\item Si $K$ est un corps, une {\bf extension} $L/K$ est une $K$-algèbre $L$ qui est un corps. 
% Un morphisme d?extensions est un morphisme de K-algèbres. On d?e?nit une sous-% % %extension ou extension intermédiaire de mani`ere ?evidente. 
%On dispose aussi de la notion %de compos?ee 
% M/K d?extensions L/K et M/L. 
\item on note $[L : K] := dim_KL$ et on dit que $L/K$ est 
{\it finie} si $[L : K] $ est finie et {\it triviale} si $[L : K] = 1$. 
\item un élément de $L$ est dit {\it algébrique} sur $K$ s'il est racine d'un polynôme à coefficients dans $K$.
\item si tout élément de $L/K$ est algébrique alors $L/K$ est dite {\it algébrique}
\item $\gamma \in L$, le {\it polynôme minimal} $\polmin_{\gamma,K}$ de $\gamma$ sur $K$ est l'unique polynôme unitaire à coefficients dans $K$ tel que $\polmin_{\gamma,K}(\gamma)=0$
\item  On dit que $\gamma \in L$ est {\it séparable} sur $K$ si diff$(\polmin_{\gamma,K},x) (\gamma)\neq 0$ ; i.e. $\polmin_{\gamma,K}$ ne possède pas de racine multiple.
\item Une extension algébrique $L/K$ est 
{\it séparable} si tout $\gamma \in  L$ est séparable sur $K$. 
\item On dit que $P\in K[x ]$ se {\it décompose} sur une extension $L/K$ en produit de facteurs 
linéaires s'il existe $c\in K$ et $u_1 ,\ldots , u_d \in  L$ avec $P = c(x-u_1 ) \ldots (x-u_d )$
\item l'extension $L$ est appelée un {\it corps de décomposition} pour $P$ si, en plus, $L = K(u_1 , . . . , u_d )$ 
\item On dit aussi qu'un polynôme non-constant $P
\in K[x ]$ est {\it séparable} (ou sans racine multiple) s'il se décompose 
sur un corps de décomposition en produit de facteurs linéaires distincts (i.e. ses racines sont deux-à-deux distinctes).
\item  $\gamma \in  L$ est séparable sur K si et seulement si $\polmin_{\gamma,K}$ est séparable. 
\item Un {\it corps de rupture} pour $P \in K[x ]$ est une extension $L/K$ telle qu'il existe $\gamma \in L$ 
avec $P (\gamma) = 0$ et $L = K(\gamma)$. 
\item Si $L/K$ est une extension et $\gamma \in L$, alors $K(\gamma)$ est un {corps de rupture} pour $\min_{\gamma,k}$ sur $K$. 
\item un corps est dit {\bf parfait} si toutes ses extensions finie sont séparables
\item {\bf Théorème de l'élément primitif (Lagrange)}\\
Si $L/K$ est une extension séparable finie alors il existe $\gamma \in L$ tel que $L = K(\gamma)$ 
\item Un corps $\overline{K}$ est {\it algébriquement clos} s'il n'existe pas d'extension 
algébrique non-triviale 
de $K$. Une clôture algébrique d'un corps K est une extension algébrique $\overline{K}/K$ qui est un corps algébriquement clos.
\item   Une extension algébrique $L/K$ est dite {\it normale} si tout $P 
\in K[x ]$ irréductible avec une 
racine dans $L$ se décompose en produit de facteurs linéaires. 
\item Une extension finie est {normale} si et seulement si c'est le corps de décomposition d'un polynôme. 
\item Une extension algébrique $L/K$ est dite {\bf galoisienne} si elle est normale et séparable. 
(i.e. si pour le polynôme minimal de tout élément de $L$ se décompose 
sur $L$ en produit de facteurs linéaires distincts.)

\end{itemize}

{\bf Corps de décomposition de $f$ }\\

{\bf Hypothèses : $k$ {\bf corps parfait} et $f$ séparable} (i.e. sans racine multiple)\\

Une {\it extension  galoisienne de $k$} est le corps des racines (de décomposition) d'un polynôme à coefficients dans $k$. 

\begin{itemize}
%\item 
\item {\bf Corps des racines et idéal maximal}

\item $\cvf$ est le corps des racines (de décomposition) de $f$ : corps des fractions de l'anneau $k[\racf]$, l'ensemble des combinaisons linéaires finies sur $k$ des $\racf^{\uplet{i}}$ où $\uplet{i} \in \N^n$ ; c'est une extension galoisienne
% $$k[\racf]:=\{\sum \lambda_i \racf^{\uplet{i}} \mid \; \uplet{i} \in \N^n, \lambda_i \in \k \}$$
\item morphisme surjectif d'évaluation :
 \begin{eqnarray*}
       \avx \longrightarrow k[\racf] , \quad p \mapsto p(\racf)
   \end{eqnarray*}   
   de noyau $\MFM$, idéal maximal.   
   \item   corps  $\avx/\MFM$ est isomorphe à $k[\racf]$ ; donc
   $$ k(\racf)= k[\racf] \simeq \avx/\MFM$$
   \item
\begin{eqnarray}
dim_k\cvf =\dim_k(\avx/\MFM)= \Card(\Grac) 
\end{eqnarray}
\item{\bf $\Gf$, Groupe de Galois de $f$}
   \item $\Gal_k(f):=\Aut_k(k(\racf))$ groupe des automorphismes de $k(\racf)$ (ensemble des $k$-endomorphismes bijectifs de  $k(\racf)$) ; $\phi \in \Gal_k(f)$ est entièrement déterminé par une bijection de
   $\{\nracf \}$ dans lui-même.
\item isomorphisme de groupes 
$$
\Gal_k(\racf) \longrightarrow \Gal_k(f), \quad \sigma \mapsto \phi_\sigma 
$$
où pour $i\in\intervalle{1}{n}\quad \phi_\sigma(\alpha_i)= \alpha_{\sigma(i)}$
%\begin{eqnarray*}
 % \phi_\sigma : \cvf \longrightarrow \cvf , \alpha_i \rightarrow \alpha_{\sigma(i)}$$
\item $\sigma \in \Gal_k(\racf)$, $\gamma \in \cvf$, $\gamma^\sigma:=\phi_\sigma(\gamma)$
\item {\bf Attention ! } $\sigma \not \in \Gal_k(\racf)$ ;\\
(i) la notation $\gamma^\sigma$ n'a aucun sens \\
(ii) $\gamma:=\sum_{\uplet{i}} \lambda_{\uplet{i}} \racf^{\uplet{i}} \in \cvf$ ; l'égalité $\gamma=0$ ne peut en aucun cas impliquer que $\sum_{\uplet{i} \in E} \lambda_{\uplet{i}} (\sigma * \racf)^{\uplet{i}}$ soit nul ; c'est justement au groupe de Galois qu'appartient le pouvoir d'assurer cette implication pour ses éléments ; il en va de même pour $\gamma = \gamma^\prime$ puisque $\gamma - \gamma^\prime=0$.

 \item {\bf Corps de rupture et polynôme minimal}
 \item le {\it polynôme minimal de $\gamma \in \cvf$ sur $k$} est  le polynôme, noté
 $\polmin_{\gamma,k}$,  irréductible sur $k$ dont $\gamma$ est racine
 \item Corps de rupture de $\gamma \in \cvf$ :
 $$k(\gamma)=k[\gamma] \simeq \ax/<\polmin_{\gamma,k}>$$
 \item $d:=\deg_x(\polmin_{\gamma,k})$ ; $$\gamma^0,\gamma^1,\ldots ,\gamma^{d-1}$$ est une base de $k$-ev de $k(\gamma)$ ;
 $$\dim_k(k(\gamma)) = d$$
\item les {\it conjugués} de $\gamma$ sur $k$ sont les racines de son polynôme minimal sur $k$.
\item les conjugués de $\gamma$ sur $k$ sont les éléments la $\Grac$-orbite de 
$\gamma$ :
$$\polmin_{\gamma,k} = \prod_{\theta \in \{\gamma^\sigma \mid \; \sigma \in \Grac \}}(x- \theta)$$
 
\item{\bf Sous-Groupes de $\Gf$ et sous-corps de $\cvf$}
\item $H<\Gal_k(\racf)$, $\cvf^H:=\{\gamma \in \cvf \mid \; (\forall \gamma \in H), \gamma^\tau = \gamma\}$
\item $H<\Gal_k(f)$, $\cvf^H:=\{\gamma \in \cvf \mid \; (\forall \phi \in H), \phi(\gamma) = \gamma\}$
\item (Evariste Galois) Soit $\gamma \in\cvf $ ; alors 
 $$(\forall \tau \in \Gal_k(\racf)), \gamma^\tau = \gamma \Leftrightarrow \gamma \in k$$
 dit autrement : $$\cvf^\Gf = k$$
 \item $$\cvf^{I_n} = \cvf$$
\item {\bf Correspondance galoisienne (Artin)}\\
% $\textmd{U}$; $\Big\{u\}, \big\{u\}, \textsf{A}, \mathrm{K}$
(i)  $K$ un corps ;
 $$k \subset K \subset \cvf \Rightarrow (\exists H\subset \Gal_k(f)), \quad K=\cvf^H$$
(ii) $H \subset \Gal_k(f) \Rightarrow k \subset \cvf^H \subset \cvf $\\

cette correspondance décrit une bijection entre les sous-groupes du groupe de Galois $\Gf$ et les corps intermédiaires entre $k$ et $\cvf$.
\item un groupe $H$ est un sous-groupe distingué de  $\Gal_k(f)$ si et seulement si l'extension $\cvf^H$ de $k$ est galoisienne 
\end{itemize}

\vspace{0.2in}
{\bf Phénomène extension-contraction : corps et idéaux galoisiens}\\

la correspondance galoisienne inhérente aux idéaux galoisiens porte sur les {\bf sur-ensembles} (et donc sur les sur-groupes) du groupe de Galois $Gal_k(\racf)$ alors que la correspondance galoisienne classique inhérente aux corps (Artin) porte sur les {\bf sous-groupes} de $Gal_k(\racf)$ (ou bien $\Gal_k(f)$)~; \\
soit 
$$H < Gal_k(\racf) < G \quad ;$$
nous avons\\
(i) $H$ décrit un corps intermédiaire entre $k$ et $\cvf$ tandis que sur les idéaux
$$I_\racf^H=\MFM  =I_\racf^{Gal_k(\racf)}$$ 
(ii) soit $K:=k(e_1,\ldots ,e_n)(\varx)^G$, le corps invariant par $G$ (voir plus bas) ; $G$ décrit un idéal intermédiaire entre $\MFS$ et $\MFM$ tandis que sur les corps
$${\widetilde K}=k=\cvf^{Gal_k(\racf)}$$
(le $\tilde \ $ symbolise la spécialisation qui envoie $x_i$ sur $\alpha_i$). \\

C'est-à-dire que pour les idéaux, les sous-ensembles de $Gal_k(\racf)$ sont associés au même idéal, i.e. à $\MFM$, alors que pour les corps, les sur-groupes de $Gal_k(\racf)$ sont associés au même corps, i.e. à $k$.

Nous pouvons voir cela encore autrement avec les $k$-algèbres $A_{G}: =\avx/I_\racf^G$.
Posons ${G_\racf} :=\Grac$ :

%\large
\begin{eqnarray}
k=\cvf^{G_\racf}=  \cvf^{G_\racf \cap G}\subset \cvf^H \subset   \cvf^{I_n}&=&\cvf\\
&\simeq&  A_{G_{\racf}}=A_{H} \subset A_{G}\subset A_{\SY_n}  
\nonumber
\end{eqnarray}
%\Large
\vspace{0.1in}
\begin{itemize}

\item {\bf extensions intermédiaires}
%$K_1,K_2$ 2 corps
\item Notation : $K_2 < K_1$ signifie $K_1/K_2$,  le corps $K_1$ est une {extension de $K_2$} ; i.e. $K_1,K_2$ corps et $K_2\subset K_1$ 
\item $k < K_2 < K_1 < \cvf$ le {\it degré} $[K_1 : K_2]$ de $K_1$ sur $K_2$ est sa dimension en tant que $K_2$-espace vectoriel (rappel)
\item $H_1<H_2<\Gf$ ; l'indice de $H_1$ dans $H_2$ et le degré de $\cvf^{H_1}$ sur $\cvf^{H_2}$ sont identiques :
$$
[H_2 : H_1] == [K_1 : K_2]
$$
\item extensions $K_3<K_2<K_1$ 
$$
[K_1 : K_3] = [K_1:K_2][K_2 : K_3]
$$

%$k < K_2 < K_1 < \cvf$, il existe un élément $K_2$-primitif du corps $K_1$.
  \item Soit $k < K_2 < K_1 < \cvf$ ; $\gamma$ est un élément {\it $K_2$-primitif de $K_1$} si $K_1=K_2(\gamma)$ 
  \item $\gamma$  est un élément  $K_2$-primitif de $K_1$ si et seulement si
  $$ \deg_x(\polmin_{\gamma,k}) = [K1 : K2] $$
  \end{itemize}

{\bf Polynôme générique }

$$pol(\varx):=\prod_{i=1}^n(x-x_i)=x^n-e_1x^{n-1}+\cdots +(-1)^ne_n$$

\begin{itemize}
\item ${\mathcal K}:=k(e_1,\ldots ,e_n)$ est le corps des coefficients de $pol(\varx)$
\item corps des racines de $pol(\varx)$ : ${\mathcal K}(\varx)=\avx$
\item $\Gal_{{\mathcal K}}(pol(\varx))= \SY_n$ (penser correspondance galoisienne)
\item ${\mathcal K}(\varx)^{\SY_n}={\mathcal K}$ ; i.e. tout polynôme symétrique s'exprime en les fonction symétriques élémentaires (théorème fondamental des fonctions symétriques)
\item $L < \SY_n$, $\Theta \in \avx$ ; {\it résolvante générique $L$-relative par $\Theta$}
$$
R_{\Theta,L} := \prod_{\Psi \in L.\Theta}(x-\Psi)
$$
\item $R_{\Theta,L}$ est le polynôme minimal de $\Theta$ sur ${\mathcal K}(\varx)^L$
\item $H < L < \SY_n$ ;
$\Theta$ est un élément ${\mathcal K}(\varx)^L$-primitif du corps ${\mathcal K}(\varx)^H$ si et seulement si $\Theta$ est un $H$-invariant $L$-primitif (d'où la terminologie !)
\item la $H$-résolvante générique $L$-relative $R_{\Theta,L}$ d'un $H$-invariant $L$-primitif $\Theta$ :
$$
 R_{\Theta,L,\varx} =\prod_{\tau \in L/H}(x-\tau.\Theta)
$$
\item le degré de l'extension ${\mathcal K}(\varx)^H$  du corps ${\mathcal K}(\varx)^L$ et l'indice de $H$ dans $L$ sont identiques :
$$
[ {\mathcal K}(\varx)^L:{\mathcal K}(\varx)^H] == [L : H]
$$
\item {\bf Extension-contraction : corps et idéaux galoisiens} \\
(i) $\MFS$ est le seul idéal galoisien définit avec $\varx$ ; i.e. tout sous-groupe de $\SY_n$ définit le même idéal galoisien, celui des relations symétriques entre les $x_i$. \\
(ii) il existe une bijection entre les sous-groupes de $\SY_n$ et les corps intermédiaires entre $k$ et $\cvx$ 
\item lien avec l'idéal des relations symétriques $\MFS$ :
$$R_{\Theta,\SY_n}==R_{\Theta,\MFS}$$
%\item Attention : comme $I_\varx^L=I_\varx^{\SY_n}=\MFS$ si $L<\SY_n$, $R_{\Theta,L}$ % ne peut s'écrire sous la forme $R_{\Theta,I}$ où $I$ est un od
\item Attention : la notation $R_{\Theta,I}:=R_{\Theta,L,\racf}$, $I$ idéal galoisien, n'est valide que pour $L=\Max(I,\racf)$ ; donc ici, elle n'a de sens que pour $L=\SY_n$.
\end{itemize}

{\bf Evaluation : point de vue lagrangien}\\

Posons $G=\Grac$.
\begin{itemize}
\item nous notons avec un $\tilde{\ }$ l'évaluation envoyant $x_i$ sur $\alpha_i$
\item $H < \SY_n$, $K:={\mathcal K}(\varx)^H$ ;
$${\widetilde K }= \cvf^{H \cap G}$$
\item en particulier, \\
(i) $(\forall H>G), \quad {\widetilde K }= \cvf^{G}$\\
(ii) $(\forall H<G), \quad {\widetilde K }= \cvf^{H}$\\
\item Evaluation de la résolvante générique :
$$\widetilde{R_{\Theta,L}}=R_{\Theta,L,\racf}$$
\item $ H < L <\SY_n$, $\Theta$ $H$-invariant $L$-primitif, $\theta={\widetilde \Theta}$ ; si $\theta$ est une racine simple de $R_{\Theta,L,\racf}$ alors $\theta$ est un élément 
$\cvf^{L\cap G}$-primitif du corps $\cvf^{H \cap G}$ de polynôme minimal $h$ sur $\cvf^{L\cap G}$ :
$$\cvf^{H \cap G} = \cvf^{L\cap G}(\theta)$$
et, par conséquent, $\deg_x(h)=[L\cap G : G\cap H]$
\item C.P. si $G<L$ alors $k=\cvf^{L\cap G}$, $R_{\Theta,L,\racf} \in \ax$ et 
$$\deg_x(h)=[G : G\cap H]$$
\item (effectivité du théorème de l'élément primitif) Soient $H_1\subset H_2 \subset \Grac$, $K_i=:\cvf^{H_i}$ ; $\Theta$ un $H_1$-invariant $H_2$-primitif $\racf$-séparable ; alors $\widetilde{\Theta}$ est un élément $K_2$-primitif de $K_1$  : 
                      $$K_1=K_2(\widetilde{\Theta})$$
\item Soient $G$ et $H$ deux sous-groupes de $S_n$ contenus dans $L$ et tels que
$L=GL=LH$ (si $L$ est un groupe cela signifie que $G$ et $H$ sont deux sous-groupes
de $L$) ;
$L=\tau_1H+\cdots +\tau_eH$ ; $H_i:=H^{\tau_i}$ ; $\Theta_i:=\tau_i\Theta$ est un $H_i$-invariant $L$-primitif ; $\theta_i:=\widetilde{\Theta_i}$ ; 
soient 
$\theta_1,\ldots ,\theta_r$ les $s$ racines d'un facteur simple $h$ de $R_{\Theta,L,\racf}$ ;
où $V:=\bigcap_{i=1}^r H_i$ ; alors 
$$
\Gal_k(h) = G/G\cap V ;
$$
pour $1\leq i\leq r$, $\deg_x(h)=[G : G\cap H_i]$
\item les groupes de Galois des facteurs d'une résolvantes ainsi que leurs degrés respectifs sont obtenus 
par le théorème précédent ; ils coïncident avec les matrices de groupes et de partitions de $L$ (voir \cite{Valibouze:95})

\item Pour les cas des {\bf racines non simples et des facteurs multiples de résolvantes, les théorèmes se généralisent} : voir \cite{VALIBOUZE:2009:HAL-00421725:1} qui les illustre avec des exemples concrets.

\end{itemize}
%\cite{ArnaudiesValibouze:97}
%\cite{AbdeljaouadSargov:02}
%\cite{MR2141285}
%\cite{Valibouze:08} 
%\cite{MR1705856}
\bibliographystyle{plain}
% je tente de mettre la dernière version dans Galois.bib
%\bibliography{/Users/avb/Bibliographie/Galois.bib}
%\bibliography{/Users/avb/Bibliographie/hal4bab5ebdc3594.bib}

\end{document}